# Ising superconductivity and anomalous metallic states in a bulk crystal with artificial unidirectional stacking layers

Xiangqi Liu, [1,†] Chen Xu, [1,†] Haonan Wang,[2,†] Runfeng Zhang,[3,†] Ziyi Zhu,[4] Zhengyang Li,[1] Lianbing Wen,[5] Ze Yan,[1] Fanbo Shen,[1] Jiawei Luo,[1,6] Zhengtai Liu,[6] Xia Wang,[1,7] Leiming Chen,[4] Ke, Qu,[2] Jianping Sun,[8] Jinguang Cheng,[8] Shiwei Wu,[5,9,*] Zhenzhong Yang,[2,*] Dawei Shen,[3,*] and Yanfeng Guo[1,10,*]

[1]State Key Laboratory of Quantum Functional Materials, School of Physical Science and Technology, ShanghaiTech University, Shanghai 201210, China

[2]Key Laboratory of Polar Materials and Devices (MOE), Department of Electronics, School of Information and Electronic Engineering, East China Normal University, Shanghai 200241, China

[3]National Synchrotron Radiation Laboratory and School of Nuclear Science and Technology, University of Science and Technology of China, Hefei 230026, China

[4]Henan Key Laboratory of Aeronautical Materials and Application Technology, Zhengzhou University of Aeronautics, Zhengzhou, Henan 450046, China

[5]State Key Laboratory of Surface Physics, Key Laboratory of Micro and Nano Photonic Structures (MOE), and Department of Physics, Fudan University, Shanghai 200433, China

[6]Shanghai Synchrotron Radiation Facility, Shanghai Advanced Research Institute, Chinese Academy of Sciences, 201204 Shanghai, China

[7]Analytical Instrumentation Center, School of Physical Science and Technology, ShanghaiTech University, Shanghai 201210, China

[8]Beijing National Laboratory for Condensed Matter Physics and Institute of Physics, Chinese Academy of Sciences, Beijing 100190, China

[9]Shanghai Research Center for Quantum Sciences, Shanghai, China

[10]ShanghaiTech Laboratory for Topological Physics, ShanghaiTech University, Shanghai 201210, China

**The two-dimensional (2D) limit in macroscopic bulk crystals provides a powerful platform for exploring exotic quantum phases. Here, we report the synthesis of a $Sr_{0.75}ClNbS_2$ superconductor that achieves unidirectional, parallel AA stacking—a configuration never before realized in a bulk crystal. Unlike conventional intercalation, which merely expands the interlayer spacing, our approach employs a planar Sr–Cl network to enforce a complete stacking reorganization, driving all $NbS_2$ layers from the native antiparallel AB stacking into a unidirectional, parallel AA arrangement. This stacking switch globally breaks inversion symmetry, transforming centrosymmetric 2H-$NbS_2$ into a noncentrosymmetric bulk crystal with $D_{3h}$ point group symmetry. Crucially, this structural design reproduces, in three dimensions, the electronic environment of an isolated monolayer, thereby preventing cancellation of the local Ising fields. As a result, strong Ising spin–orbit coupling and spin-split bands persist throughout the bulk. Transport measurements reveal extreme superconducting anisotropy ($\gamma$ ~ 77), an in-plane upper critical field ($B_{c2}^{//}$ ~ 10.65 T) that far exceeds the Pauli paramagnetic limit, and clean-limit superconductivity indicative of high crystalline quality. Moreover, magnetotransport uncovers a novel magnetic-field-induced anomalous metallic state characterized by finite dissipation yet a vanishing Hall response. Direct band-structure measurements corroborate the layer-decoupled, quasi-2D electronic nature of the system. This work establishes stacking-geometry engineering as a powerful strategy to artificially enforce a globally noncentrosymmetric, quasi-2D superconducting state in bulk crystals, paving the way for designing quantum materials with tunable crystalline symmetry and electronic band topology.**

[†]These authors contributed equally to this work.

Xiangqi Liu, Chen Xu, Haonan Wang and Runfeng Zhang.

[*]Correspondence:

swwu@fudan.edu.cn, zzyang@phy.ecnu.edu.cn, dwshen@ustc.edu.cn,

guoyf@shanghaitech.edu.cn.

## Introduction

Monolayer transition-metal dichalcogenides (TMDs) in the hexagonal phase (1H) feature a trigonal prismatic coordination, where a plane of transition-metal atoms is sandwiched between two chalcogen layers. This structural motif inherently lacks in-plane inversion symmetry while preserving out-of-plane mirror symmetry. When combined with the strong spin-orbit coupling (SOC) arising from heavy 4*d* transition-metal atoms such as Nb, this symmetry configuration imposes a distinctive constraint: the crystal electric field ($\varepsilon$) is strictly confined to the two-dimensional (2D) plane. As a result, the SOC lifts the spin degeneracy at finite momentum $\boldsymbol{k}$ without requiring an external magnetic field, giving rise to a momentum-dependent effective Zeeman field $\boldsymbol{H}_{\mathrm{SO}}(\boldsymbol{k}) \propto \boldsymbol{k} \times \boldsymbol{\varepsilon}$. Since electronic motion is confined to the plane, this effective field points strictly out-of-plane, locking the electron spins along the out-of-plane direction. At the K and K' valleys, electrons with opposite momenta exhibit opposite spin orientations, thereby realizing robust Ising-type spin-momentum locking[1-10]. Notably, such locking is typically absent in bulk 2H-TMDs, where antiparallel stacking between adjacent layers restores global inversion symmetry and spin degeneracy.

To experimentally access these effects, various techniques have been developed to isolate or synthesize TMD monolayers. The most direct methods include mechanical exfoliation and the growth of continuous monolayer films via chemical vapor deposition (CVD) or molecular beam epitaxy (MBE)[11-13]. Previous studies have demonstrated the CVD growth of $MoS_2$ multilayers with either mixed AA/AB or pure AA stacking. These distinct stacking configurations can give rise to unusual physical properties, such as the intense second-harmonic generation (SHG) signals[14,15]. While these top-down and bottom-up approaches have provided essential platforms for investigating Ising-type interactions, several critical limitations persist. First, the lateral dimensions of exfoliated or grown monolayer samples are typically limited to the micrometer scale, which restricts experimental probes primarily to electrical transport. Macroscopic thermodynamic and magnetic measurements, such as specific heat, thermal conductivity, and bulk magnetization, remain exceedingly challenging. Second, unprotected TMD monolayers are intrinsically fragile and highly susceptible to environmental degradation. Although encapsulation with hexagonal boron nitride flakes can partially mitigate this issue, devices often still suffer from long-term instability upon extended exposure. Consequently, the measured superconducting properties frequently exhibit considerable variability across different device batches. These challenges motivate the synthesis of stable, macroscopic bulk crystals that retain both strong Ising SOC and broken inversion symmetry, offering a highly desirable platform for systematically investigating noncentrosymmetric superconductivity.

Chemical intercalation provides a powerful route for modulating interlayer spacing and Josephson coupling in bulk TMDs. By inserting insulating spacer layers between TMD sheets, interlayer electron hopping can be substantially suppressed, realizing a weak-coupling quasi-2D limit that is ideal for testing theoretical models of Ising superconductivity[16-22]. Nevertheless, despite successful intercalation of various spacer polytypes into bulk TMD single crystals, a fundamental structural constraint persists. In most intercalated compounds, the H-$MX_2$ layers retain the antiparallel (ABA) stacking sequence inherited from the parent 2H phase. As adjacent layers act as inversion partners, their effective Ising fields $\boldsymbol{H}_{SO}$ at a given valley point in opposite directions. Consequently, any residual, nonzero Josephson coupling between layers partially cancels the net Ising protection.

In this work, we overcome this limitation by designing and constructing an ideal "monolayer-like" platform within a bulk crystal, achieved through enforcing perfectly parallel stacking of H-$NbS_2$ layers with extremely weak interlayer coupling. Specifically, we employ a planar Sr-Cl network as an intercalating framework to lock the H-$NbS_2$ monolayers into an exact AA stacking sequence. This unique structural arrangement intrinsically prevents cancellation of the Ising field, enabling us to probe robust Ising-type SOC alongside 2D superconductivity and anomalous metallic behavior under extreme magnetic fields. By faithfully emulating the ideal monolayer limit, our approach preserves both the noncentrosymmetric crystal structure and exceptionally strong Ising SOC within a stable macroscopic bulk material. This establishes a new platform for investigating fascinating Ising superconductivity as well as a range of other rich electrical and optical properties.

The details for single crystal growth, quality examinations, energy dispersive spectroscopy (EDS) characterizations, magneto-transport, atomic-scale characterizations, nonlinear optical second-harmonic generation (SHG) and angle-resolved photoemission spectroscopy (ARPES) measurements of $Sr_{0.75}ClNbS_2$ are presented in the Supplemental Material (SM) which includes Figs. S1-S11 and Refs. [5, 37-43]. For comparison, $Ba_{0.75}ClNbS_2$ crystals were also grown, and the details of their growth and basic characterization are provided in the SM.

## Results and Discussions

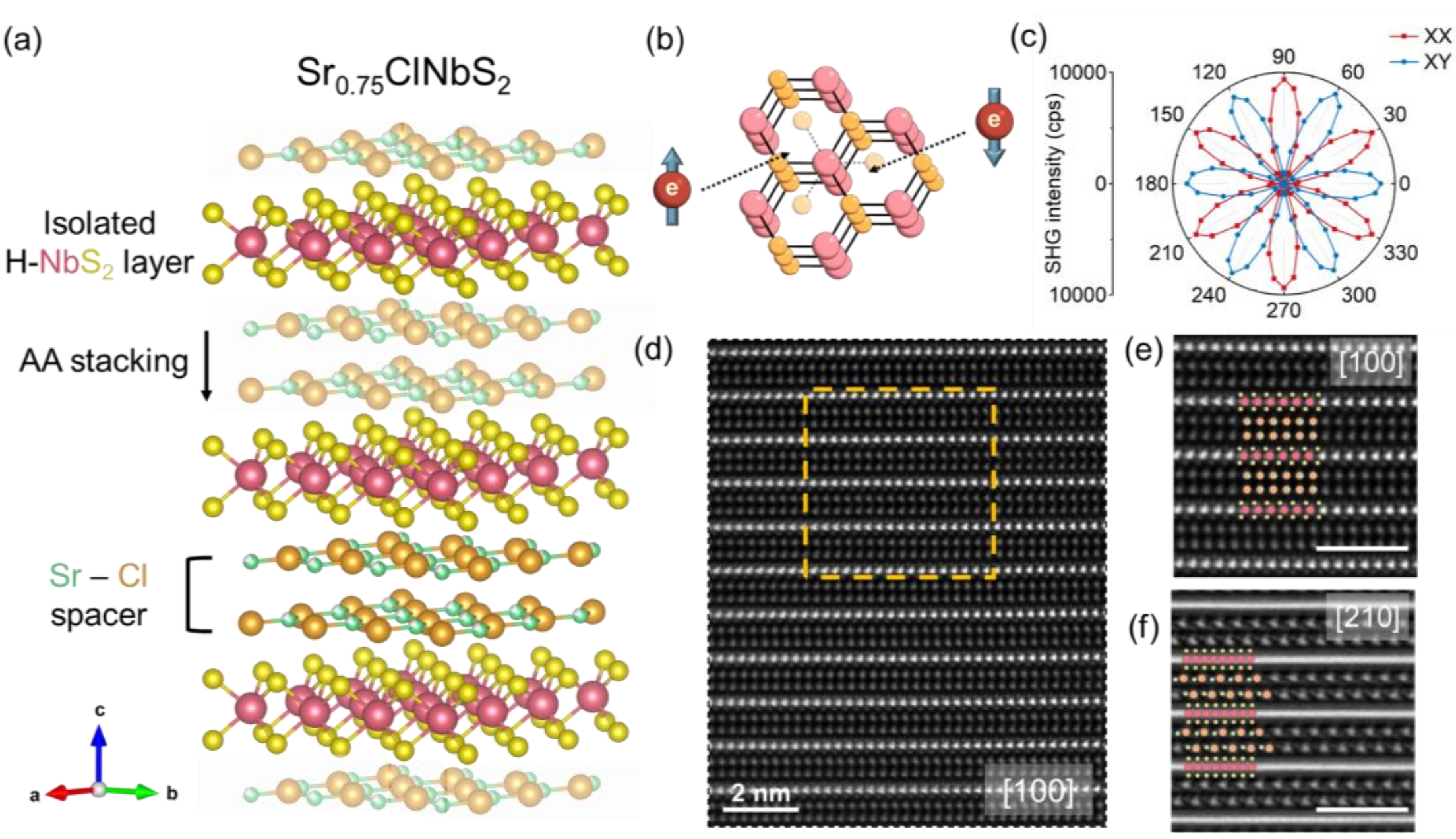

**Fig. 1.** (a) Crystal structure of $Sr_{0.75}ClNbS_2$, highlighting perfectly parallel AA stacking of H-$NbS_2$ layers separated by Sr-Cl spacer layers. (b) Schematic illustration of the out-of-plane spin-momentum locking effect in AA-stacked H-$NbS_2$ layers, which is strictly protected by global inversion symmetry breaking. (c) Nonlinear optical SHG patterns collected under parallel and perpendicular polarizations. (d, e) Cross-sectional HAADF-STEM images viewed along the [100]

zone axis of the lattice, overlaid with the refined structural model. Due to closely overlapping projected positions along this orientation, the Sr and Cl atoms cannot be individually resolved. (f) Cross-sectional HAADF-STEM image acquired along the [210] zone axis, where the Sr and Cl atomic columns within the spacer layer are clearly resolved.

The crystal structure of $Sr_{0.75}ClNbS_2$ was determined by single-crystal X-ray diffraction (SXRD) at room temperature (Fig. S1). As shown in Fig. 1(a), the structure adopts an AA stacking of H-$NbS_2$ layers separated by Sr-Cl spacer layers along the crystallographic *c*-axis. Within the H-type TMD layers, each Nb atom is trigonal-prismatically coordinated by six S atoms. The in-plane symmetry and lattice mismatch were further probed by SXRD patterns collected on the (*hk0*) plane (Fig. S1), where two distinct sets of Bragg reflections are clearly resolved: one from the hexagonal $NbS_2$ layers (yellow dashed lines, first Brillouin zone) and the other from the monoclinic Sr-Cl spacer layers (orange dashed lines). Their clear separation confirms the mutual incommensurability between the in-plane lattices of the two subsystems, from which we extracted lattice parameters of $a = b = 3.35$ Å for the $NbS_2$ layer and $a = 3.35$ Å, $b = 2.98$ Å for the Sr-Cl layer. Interestingly, the stacking sequence can be modulated by substituting Sr with Ba: the Ba-Cl framework favors a parallel ABCA stacking of H-$NbS_2$ layers, giving rise to a rhombohedral 3R phase. The extracted in-plane lattice parameters for the $NbS_2$ layer ($a = b = 3.35$ Å) and the Ba-Cl spacer ($a = 3.35$ Å, $b = 2.91$ Å) are highly consistent with those observed in the Sr-Cl counterpart.

In contrast to the 2H-$NbS_2$, adjacent layers in $Sr_{0.75}ClNbS_2$ act as symmetry-equivalent partners rather than inversion partners, so their effective Ising fields $\boldsymbol{H}_{SO}$ at a given valley point in the same direction. Consequently, the net Ising protection is preserved, illustrated in Fig. 1(b). To probe the quasi-monolayer physical properties arising from the unique AA stacking mode, we measured the nonlinear optical SHG response of $Sr_{0.75}ClNbS_2$. As shown in Fig. 1(c), the SHG signal exhibits a striking sixfold symmetry pattern that strictly follows $\sin^2(3\theta)$ and $\cos^2(3\theta)$ dependencies under parallel and perpendicular polarizations, respectively. This observation provides unambiguous evidence of global inversion symmetry breaking, consistent with the $D_{3h}$ point group and mimicking the behavior of a standalone H-$NbS_2$ monolayer[23,24]. Notably, such symmetry reduction is completely absent in centrosymmetric bulk 2H-$NbS_2$ ($D_{6h}$ symmetry) and even-layered flakes ($D_{3d}$ symmetry). Furthermore, the uniform SHG intensity distribution corroborates the unidirectional nature of the AA stacking alignment throughout the bulk crystal.

The cross-sectional structure of $Sr_{0.75}ClNbS_2$ was first examined along the [100] zone axis using high-angle annular dark-field scanning transmission electron microscopy (HAADF-STEM), with the refined structural model overlaid (Figs. 1(d) and 1(e)). Consistent with SXRD results, the unit cell consists of H-$NbS_2$ layers alternating with Sr-Cl spacer layers. Notably, the H-$NbS_2$ layers stack in perfect parallel registry, resulting in a substantially reduced $c$-axis lattice parameter ($c$ = 11.4 Å) relative to other common intercalated TMDs. However, along the [100] direction, the Sr and Cl atoms occupy closely spaced Wyckoff positions, rendering them difficult to resolve individually. To overcome this limitation, we imaged the same material along the [210] zone axis (Fig. 1(f)). From this perspective, the Sr atoms form a zigzag chain-like alternating arrangement between adjacent $NbS_2$ layers, while the Cl atoms become clearly distinguishable at their crystallographic sites.

To investigate the superconductivity arising from the unique AA-stacking configuration, we performed electrical transport measurements on bulk $Sr_{0.75}ClNbS_2$ single crystals. Fig. 2(a) shows the temperature dependence of normalized resistance at low temperatures. The onset superconducting transition temperature ($T_{c\text{-onset}}$) and the zero-resistance temperature ($T_{c\text{-zero}}$) are defined by the intersections of the extrapolated transition region with the normal-state and zero-resistance baselines, respectively. Compared to pristine 2H-$NbS_2$ ($T_c$ ~ 5.5 K), Sr-Cl intercalation reduces the critical temperature to $T_{c\text{-onset}}$ = 1.54 K and $T_{c\text{-zero}}$ = 0.88 K. In contrast, temperature-dependent resistance measurements on $Ba_{0.75}ClNbS_2$ (Fig. S7) reveal a much lower superconducting transition ($T_c$ < 400 mK), implying that the AA stacking sequence, rather than the mere expansion of interlayer spacing, plays a dominant role in generating the superconductivity. Figs. 2(b) and 2(c) present the temperature-dependent magnetoresistance under magnetic fields applied perpendicular and parallel to the *ab*-plane, respectively. For out-of-plane fields, superconductivity is suppressed predominantly by the orbital effect, which induces vortex motion and core overlap, eventually destroying macroscopic phase coherence. In contrast, under in-plane fields, a strong effective $\boldsymbol{H}_{SO}$ within the individual H-$NbS_2$ layers firmly pins the electron spins along the out-of-plane direction. This robust spin-momentum locking mechanism effectively competes with the Pauli paramagnetic pair-breaking effect[2,3,25], resulting in a broadened resistive transition before the system fully enters the normal state.

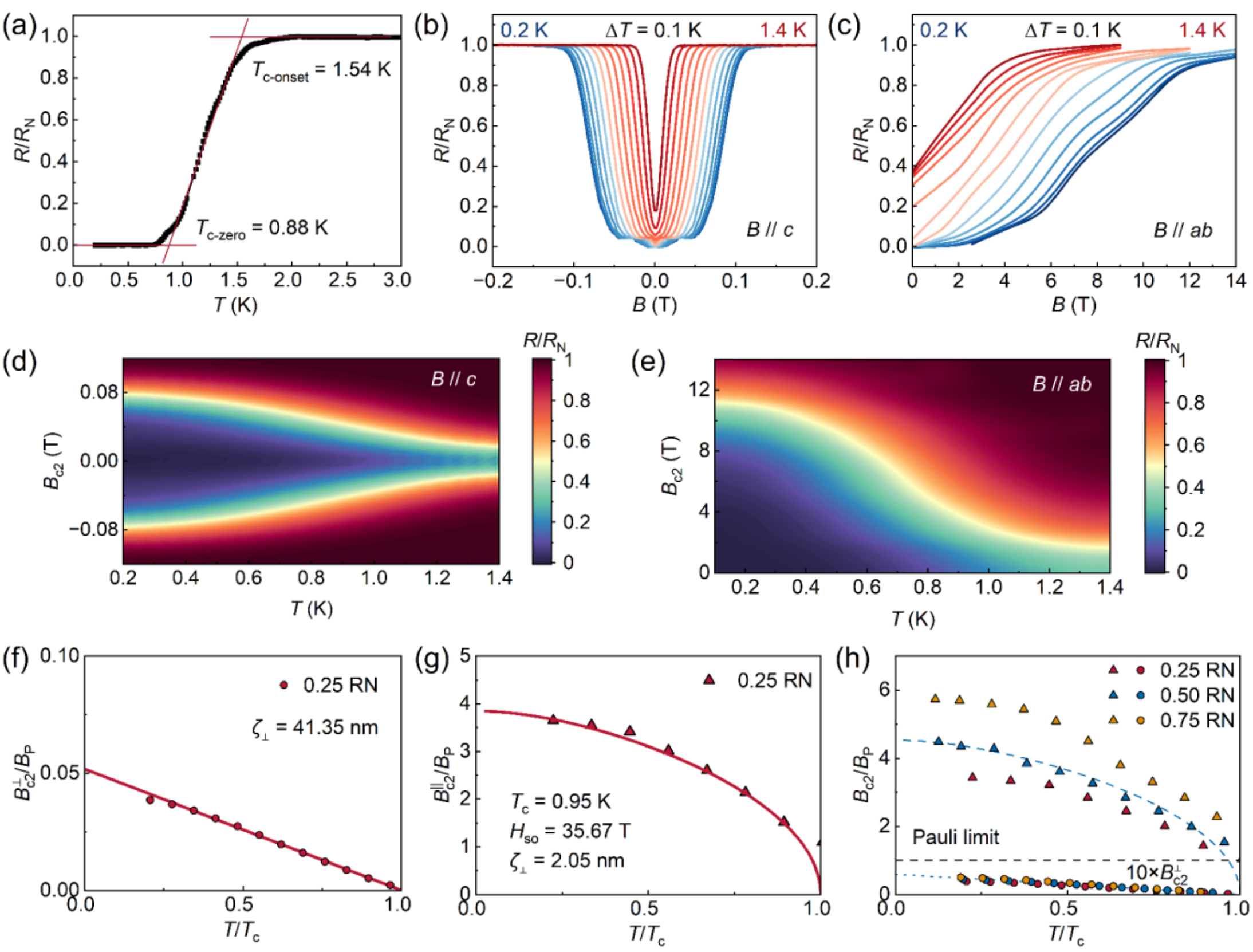

**Fig. 2.** (a) Temperature dependence of the normalized longitudinal resistance at low temperatures (< 3 K). (b, c) Magnetic-field dependence of the normalized resistance measured at temperatures from 0.2 K to 1.4 K in 0.1 K steps, with the external field applied perpendicular (b) and parallel (c) to the *ab*-plane, respectively. (d, e) Corresponding color-mapped magnetoresistance contour plots, showing the $B_{c2}$-$T$ phase diagrams for out-of-plane (d) and in-plane (e) field orientations. (f) Temperature dependence of the out-of-plane upper critical field $B_{c2}^{\perp}(T)$. The linear dependence is consistent with the conventional 3D GL orbital depairing mechanism. (g) Temperature dependence of the in-plane upper critical field $B_{c2}^{//}(T)$. The data are well described by the pair-breaking model incorporating spin-momentum locking, providing strong evidence for bulk Ising superconductivity in $Sr_{0.75}ClNbS_2$. (h) Comparison of the in-plane and out-of-plane upper critical fields extracted using different normal-state resistance criteria, emphasizing the pronounced superconducting anisotropy.

From these magnetoresistance data, we constructed the corresponding $B_{c2}$-$T$ phase diagrams (Figs. 2(d) and 2(e)). For out-of-plane fields (Fig. 2(f)), $B_{c2}^{\perp}(T)$ follows a linear relationship well described by the 3D Ginzburg-Landau (GL) orbital depairing model (see SM for details). The extremely low out-of-plane upper critical field yields a remarkably large in-plane GL coherence length $\xi^{//}$ = 41.35 nm, which exceeds the in-plane lattice constant of the $NbS_2$ layer ($a$ = $b$ = 3.35 Å) by more than two orders of magnitude. For in-plane fields (Fig. 2(g)), $B_{c2}^{//}(T)$ exhibits a square-root temperature dependence near $T_c$. The experimental data are well captured by the pair-breaking model incorporating spin-momentum locking, originally developed for monolayer H-$NbSe_2$[5,26]. From this fit, we extract the effective Ising field $H_{SO}$ = 35.67 T, which is the key parameter governing the substantially enhanced in-plane upper critical field.

Based on the 2D GL description, the out-of-plane coherence length $\xi^{\perp}$ corresponds to an effective superconducting thickness $d_{SC}$ = $\xi^{\perp}$ = 2.05 nm. Although this value slightly exceeds the interlayer spacing between adjacent $NbS_2$ layers, indicating the persistence of weak residual Josephson coupling, it highlights the difficulty of completely eliminating interlayer coupling solely through intercalation. Remarkably, it is the synergistic combination of the Sr-Cl spacer layer and the designed AA-type stacking that successfully preserves local inversion symmetry breaking and the concomitant Ising spin-orbit field. As a result, comparing $B_{c2}^{\parallel}$ and $B_{c2}^{\perp}$ under various criteria (0.25 $R_N$, 0.50 $R_N$ and 0.75 $R_N$) reveals an exceptionally large superconducting anisotropy of $\gamma = B_{c2}^{\parallel}/B_{c2}^{\perp}$~77 (Fig. 3h). This giant anisotropy significantly surpasses that of most pristine and intercalated bulk 2D superconductors (for instance, the organic superconductor κ-$(BEDT\text{-}TTF)_2Cu(NCS)_2$ exhibits $\gamma$ = 220) and rivals those of genuine monolayer TMDs, such as 1H-$MoS_2$ ($\gamma$ = 83) and 1H-$TaS_2$ ($\gamma$ = 28)[7,27,28].

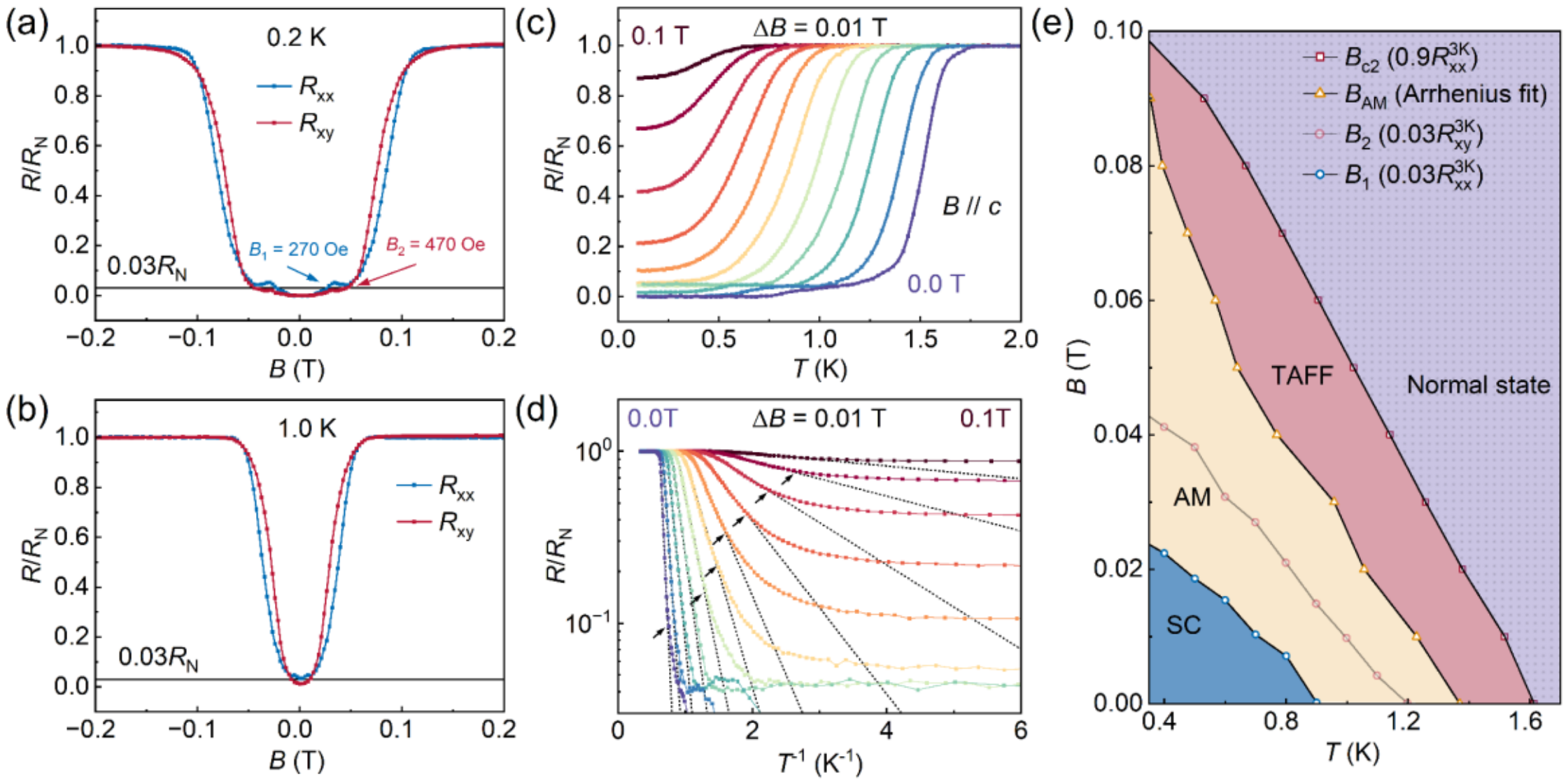

**Fig. 3.** (a, b) Magnetic-field dependence of the normalized $R_{xx}$ and $R_{xy}$ measured at 0.2 K (a) and 1.0 K (b). At 0.2 K, the emergence of a finite $R_{xx}$ plateau while $R_{xy}$ remains strictly zero provides striking evidence for a particle-hole symmetric AMS. (c) Temperature dependence of $R_{xx}$ under various out-of-plane magnetic fields. (d) Arrhenius plots of $R_{xx}(T)$ at different magnetic fields. The black arrows mark the characteristic crossover temperatures where $R_{xx}$ deviates from the linear scaling predicted by the classical TAFF model, signaling the onset of the quantum-fluctuation-dominated AMS. (e) Comprehensive *B*-*T* phase diagram delineating the boundaries of the macroscopic superconducting state, the TAFF regime, the AMS, and the normal state.

Anomalous metallic states (AMS) have attracted substantial interest in the study of 2D superconductivity, particularly in the context of magnetic-field-, structural-, and electric-doping-driven superconductor-metal and superconductor-insulator quantum phase transitions. TMDs serve as ideal platforms for exploring dimensionality-tunable AMS. As the dimensionality is reduced toward the 2D limit, enhanced quantum phase fluctuations, vortex dynamics, and strong electronic correlations give rise to a variety of thickness-dependent quantum phenomena, including the exotic Bose-metal phase[29-31]. In $Sr_{0.75}ClNbS_2$, the pronounced quasi-2D character arising from its unique AA stacking provides an exceptional bulk platform for investigating superconducting quantum criticality and AMS via electrical transport measurements. Fig. 3(a) presents the normalized longitudinal resistance ($R_{xx}$) and Hall resistance ($R_{xy}$) as functions of the perpendicular magnetic field applied at 0.1 K. The onset of normal-state transport is defined by criteria $R_{\mathrm{xx}}(B_1) = 0.03\ R_{\mathrm{xx}}^{3\mathrm{K}}$ and $R_{\mathrm{xy}}(B_2) = 0.03\ R_{\mathrm{xy}}^{3\mathrm{K}}$. Notably, in an intermediate field regime, $R_{xx}$ deviates from zero while $R_{xy}$ remains negligibly small. This distinct absence of a Hall signal alongside finite longitudinal dissipation strongly suggests preserved particle-hole symmetry — a hallmark of bosonic transport in which localized and uncondensed Cooper pairs persist even though macroscopic phase coherence is destroyed by the external field, thereby giving rise to the AMS. Upon raising the temperature to 1 K (Fig. 3(b)), both $R_{xx}$ and $R_{xy}$ emerge concurrently with increasing magnetic field, indicating that thermal fluctuations suppress the AMS.

To quantitatively analyze the AMS, the temperature-dependent $R_{xx}$ under various out-of-plane magnetic fields is plotted following the Arrhenius convention (Figs. 3(c,d)). Below a characteristic temperature threshold $T_{\mathrm{AM}}$ (indicated by the black arrow in Fig. 3(d)), the resistance curves systematically deviate from the linear scaling predicted by the thermally activated flux flow (TAFF) model[32-34]. This low-temperature resistance flattening confirms the transition into a quantum-fluctuation-driven AMS. By synthesizing the *R*-*B* and *R*-*T* transport behaviors, a comprehensive phase diagram (Fig. 3(e)) mapping the field-induced transitions is constructed. The upper boundary of the macroscopic superconducting phase is delineated by $B_1$, where $R_{xx}$ reaches 3% of its normal-state value, while the lower boundary of the normal state is defined by the upper critical field $B_{\mathrm{c2}}$ (where $R_{xx}$ reaches 90% of its normal-state value). Sandwiched between these boundaries lie the TAFF regime and the AMS. The crossover into the AMS is defined by $B_{\mathrm{AM}}$ — the boundary at which the *R*-*T* transport deviates from classical Arrhenius thermal activation (this boundary is close to $B_2$ derived from the Hall signal). The observation of a robust, magnetic-field-induced AMS is consistent with findings in strictly 2D systems such as ultrathin $NbSe_2$, $SnSe_2$, and $WS_2$, further demonstrating that the AA-stacked bulk crystal successfully emulates the quantum critical behavior of the true 2D monolayer limit[32-36].

To elucidate the electronic properties underlying the observed Ising-type superconductivity, we performed ARPES measurements on $Sr_{0.75}ClNbS_2$ single crystals. Figs. 4(a,b) present the Fermi surface mapping at the Fermi level ($E_F$) and the constant energy contour at $E_F$ - 0.2 eV, respectively. With increasing binding energy, the spectral weight of the pockets expands notably, unambiguously confirming their hole-like character. The experimental Fermi surface comprises a prominent hexagonal hole-pocket contour centered at the $\bar{\Gamma}$ point, accompanied by additional hole pockets at the $\bar{K}$ points. Fig. 4(c) displays the measured band dispersions along the high-symmetry $\bar{K}$-$\bar{\Gamma}$-$\bar{M}$-$\bar{K}$ directions. Remarkably, the overall band structure closely resembles that of a freestanding H-$NbS_2$ monolayer, consistent with prior experimental and theoretical studies on the monolayer limit[7,17,37]. To rigorously assess the dimensionality of these electronic states, we performed photon-energy-dependent ARPES measurements to map the out-of-plane $k_z$ dispersion (see SM for details). The constant-energy contours exhibit negligible dependence on incident photon energy, providing unambiguous momentum-space evidence for the highly 2D nature of the electronic structure in $Sr_{0.75}ClNbS_2$. This distinct layer-decoupled feature confirms that intercalation of Sr-Cl spacer layers effectively suppress interlayer coupling, driving the macroscopic bulk crystal deep into the quasi-2D limit.

Furthermore, compared with pristine bulk 2H-$NbS_2$, the visibly reduced area of the hole pockets in $Sr_{0.75}ClNbS_2$ implies a lower hole concentration. This observation offers direct evidence of a substantial electron-doping effect arising from charge transfer from the intercalated Sr-Cl spacer layers to the $NbS_2$ host layers. Consequently, the suppression of $T_c$ relative to pristine 2H-$NbS_2$ single crystals can be primarily attributed to a reduction in the electronic density of states at $E_F$—a direct consequence of this electron doping. Given this highly 2D electronic structure and the globally broken inversion symmetry structurally enforced by the unique AA stacking, the inherent Ising-type SOC remains robustly preserved in the bulk. This lays a solid foundation for future polarization-dependent ARPES measurements selectively targeting the inequivalent K and $\bar{K}$ valleys to further corroborate the physics of spin-valley locking.

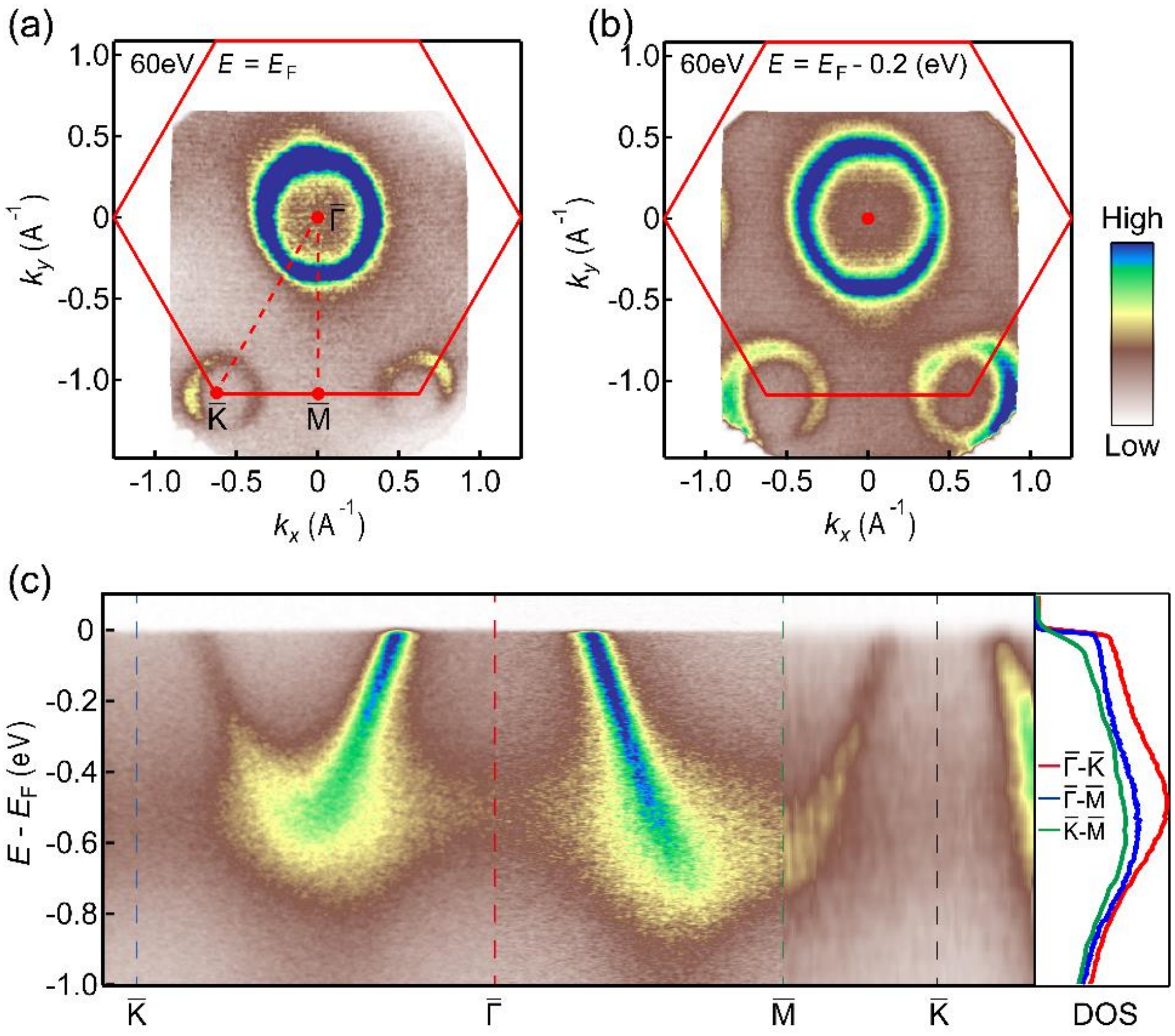

**Fig. 4.** (a) Fermi surface mapping of the $Sr_{0.75}ClNbS_2$ single crystals. (b) Constant-energy contour extracted at a binding energy of $E_F$ - 0.2eV. (c) Left: ARPES intensity plot detailing the band dispersions along the high-symmetry $\bar{K}$-$\bar{\Gamma}$-$\bar{M}$-$\bar{K}$ direction. Right: the corresponding integrated DOS along these high-symmetry paths.

## Conclusion

In summary, we have successfully engineered a stable bulk van der Waals superlattice, $Sr_{0.75}ClNbS_2$, that faithfully replicates the noncentrosymmetric electronic environment of an isolated monolayer within a macroscopic single crystal. By enforcing a strictly parallel AA-stacking sequence via planar Sr-Cl spacer layers, this system effectively eliminates the parasitic cancellation of Ising protection inherent to conventional AB-stacked misfit compounds. Global inversion symmetry breaking, unambiguously confirmed by SHG and ARPES, sustains a large out-of-plane effective

Ising field. This robust spin-momentum locking, together with an exceptionally low defect scattering rate placing the system deep in the clean limit, gives rise to giant superconducting anisotropy and a low-temperature upturn in the upper critical field—likely suggestive of an emergent orbital Fulde-Ferrell-Larkin-Ovchinnikov state[38].

Furthermore, the pristine low-dimensional confinement amplifies quantum phase fluctuations, leading to a pronounced magnetic-field-induced anomalous metallic state where particle-hole symmetry is preserved through bosonic Cooper-pair transport. Ultimately, the synthesis of $Sr_{0.75}ClNbS_2$ establishes a highly scalable and robust material platform, bridging the gap between delicate monolayer device physics and macroscopic thermodynamic studies of noncentrosymmetric superconductivity, quantum criticality, and valleytronics.

**Acknowledgements**

The authors acknowledge the National Key R&D Program of China (Grants No. 2023YFA1406100, 2024YFA1408400, 2023YFA1406304 and 2022YFA1403300) and the National Nature Science Foundation of China (Grants No. 52572129, 12522407, 12494593 and 12427807). Y.F.G. acknowledges the Science and Technology Commission of Shanghai Municipality (25DZ3008200) and open research fund of Beijing National Laboratory for Condensed Matter Physics (2023BNLCMPKF002). Z.Z.Y. acknowledges financial support from the Science and Technology Commission of Shanghai Municipality (Grant No. 24JD1401200). W.D.S. thanks the HFNL Self-Deployed Project (Grant No. ZB2602000302), Anhui Provincial Natural Science Foundation (Grant No. 2408085J003). We thank the Shanghai Synchrotron Radiation Facility (SSRF) of BL03U (31124.02.SSRF.BL03U) for the assistance of ARPES measurements. The authors also thank the Analytical Instrumentation Center (#SPST-AIC10112914) and the Double First-Class Initiative Fund of ShanghaiTech University.

# SI

## I. Crystal growth

High-quality single crystals of $Sr_{0.75}ClNbS_2$ and $Ba_{0.75}ClNbS_2$ were synthesized using a flux method. First, polycrystalline $NbS_2$ precursor powder was prepared by mixing high-purity Nb powder (Macklin, 99.99%) and S powder (Macklin, 99.99%) in a molar ratio of 1:2.1. The mixture was sealed in an evacuated quartz tube, heated to 973 K over 10 hours, and held at this temperature for 50 hours.

The resulting $NbS_2$ polycrystalline powder was then combined with $SrCl_2$ or $BaCl_2$ (Alfa Aesar, 99.99%) in a weight ratio of 1:2. The charge mixture was prepared in an inert atmosphere, sealed into quartz ampoules under a vacuum better than $10^{-4}$ Torr, and placed in a box furnace. The ampoules were heated to 1150 K, held at this temperature for a minimum of 120 hours, then cooled to 1000 K at a rate of 2 K/h, followed by furnace cooling to room temperature. Single crystals of $Sr_{0.75}ClNbS_2$ and $Ba_{0.75}ClNbS_2$ were subsequently extracted from the crucible. The obtained crystals are stable in air, exhibiting no signs of degradation or structural phase transitions under ambient conditions.

## II. Compositions and phase examinations of $Ba_{0.75}ClNbS_2$ and $Sr_{0.75}ClNbS_2$

Chemical compositions of the as-grown $Ba_{0.75}ClNbS_2$ and $Sr_{0.75}ClNbS_2$ single crystals were determined by energy-dispersive X-ray spectroscopy (EDS), with key structural and compositional features summarized in Figs. S1 and S2. For the Ba-analogue (Figs. S1(a-d)), the crystals typically grow to dimensions of $\sim 2 \times 3 \times 0.4$ mm$^3$, and their cleaved surfaces exhibit macroscopic stripe-like domains (Fig. S1(b))—likely a manifestation of structural modulation induced by the monoclinic Ba-Cl spacer layers. EDS statistical analysis (Fig. S1(c)) yields an atomic ratio close to the nominal stoichiometry, albeit with a noticeable sulfur deficiency (Nb:S ≈ 1:1.73, corresponding to ~ 13% S-site vacancies). The (*hk0*) SXRD pattern (Fig. S1(d)) clearly resolves two independent sets of Bragg reflections: one from the hexagonal $NbS_2$ sublattice (blue dashed lines) and the other from the monoclinic Ba-Cl spacer layers (orange dashed lines), confirming their mutual incommensurability.

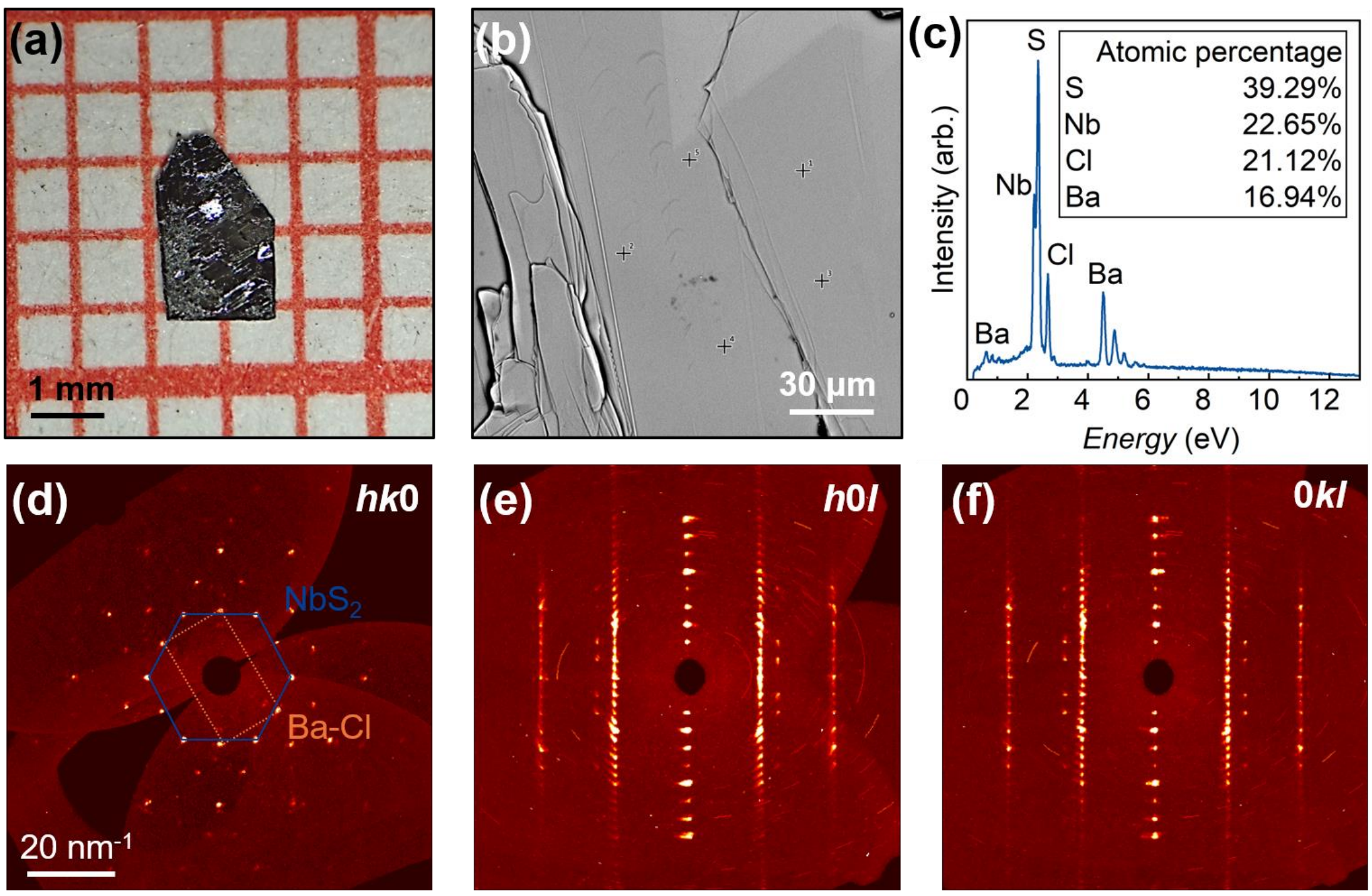

**Fig. S1.** (a) Optical image of a typical $Ba_{0.75}ClNbS_2$ single crystal. (b) SEM image of the cleaved surface of a $Ba_{0.75}ClNbS_2$ crystal, showing the characteristic layered morphology. (c) Representative EDS spectrum confirming the elemental composition with a stoichiometric molar ratio of Ba:Cl:Nb:S ≈ 0.75:1:1:1.7. (d-f) Single crystal XRD patterns in reciprocal space, viewed along the (0*kl*), (*h*0*l*), and (*hk*0) planes, respectively.

In contrast, the Sr-analogue exhibits notably larger crystal dimensions (Fig. S2(a)). EDS analysis on a clean surface (Fig. S2(b)) reveals a relatively sulfur-rich composition (Nb:S ≈ 1:2.25, Fig. S2(c)) compared with the Ba-phase. Furthermore, Figs. S2(d-f) present the SXRD patterns collected on the (*hkl*) reciprocal lattice plane. As seen in the (*hk0*) plane, two independent sets of Bragg reflections are clearly resolved: one corresponds to the hexagonal $NbS_2$ layers (yellow dashed lines, first Brillouin zone), and the other originates from the monoclinic Sr-Cl spacer layer (orange dashed lines). Their distinct separation reveals that the in-plane lattices of the two components are mutually incommensurate. Moreover, the (*hk0*) SXRD pattern of $Sr_{0.75}ClNbS_2$ (Fig. S2(d)) shows that the Bragg reflections from the Sr-Cl spacer layers display a distinct reflection symmetry (Figs. S2(e,f)), in contrast to those of the Ba-intercalated compound (Fig. S1(d)), highlighting the different structural arrangements of the two spacer frameworks.

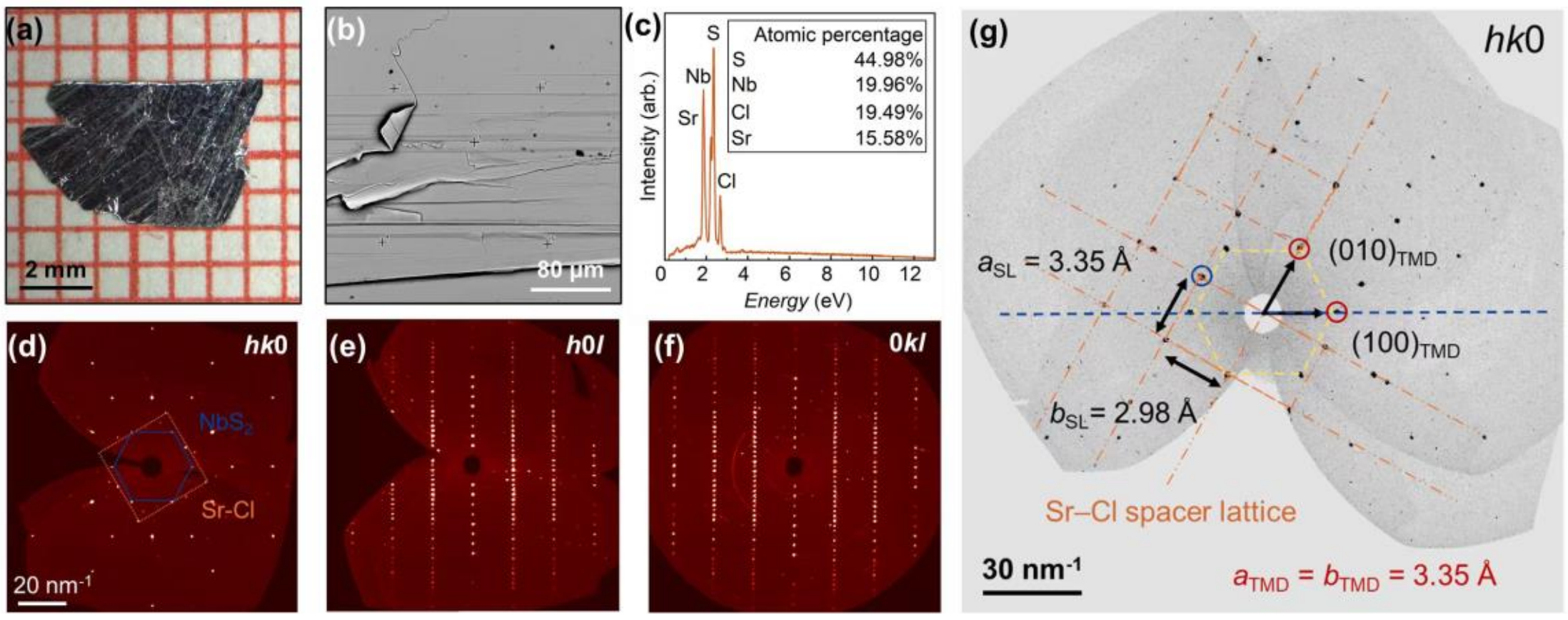

**Fig. S2.** (a) Optical image of a typical $Sr_{0.75}ClNbS_2$ single crystal. (b) SEM image of the cleaved surface of a $Sr_{0.75}ClNbS_2$ single crystal. (c) Representative EDS spectrum confirming the elemental composition with a stoichiometric molar ratio of Sr: Cl: Nb: S ≈ 0.75:1:1:2.2. (d-f) SXRD patterns in reciprocal space collected at room temperature, viewed along the (*0kl*), (*h0l*), and (*hk0*) planes, respectively. (g) displays the SXRD pattern collected on the (*hk*0) reciprocal lattice plane, revealing two distinct sets of Bragg reflections. The set enclosed by yellow dashed lines (first Brillouin zone) originates from the hexagonal H-$NbS_2$ layers, while the set marked by orange dashed lines arises from the monoclinic Sr-Cl spacer layers.

## III. CS-corrected scanning transmission electron microscopy measurements

The cross-sectional transmission electron microscopy (TEM) sample was prepared using a focused ion beam scanning electron microscopy (FIB-SEM) system, specifically a Helios G4 UX instrument from Thermo Fisher. Atomic-resolution high-angle annular dark-field (HAADF) and annular bright-field (ABF) scanning transmission electron microscopy (STEM) observations were conducted using a 300 kV spherical aberration (Cs)-corrected STEM, the JEM-ARM300F from JEOL.

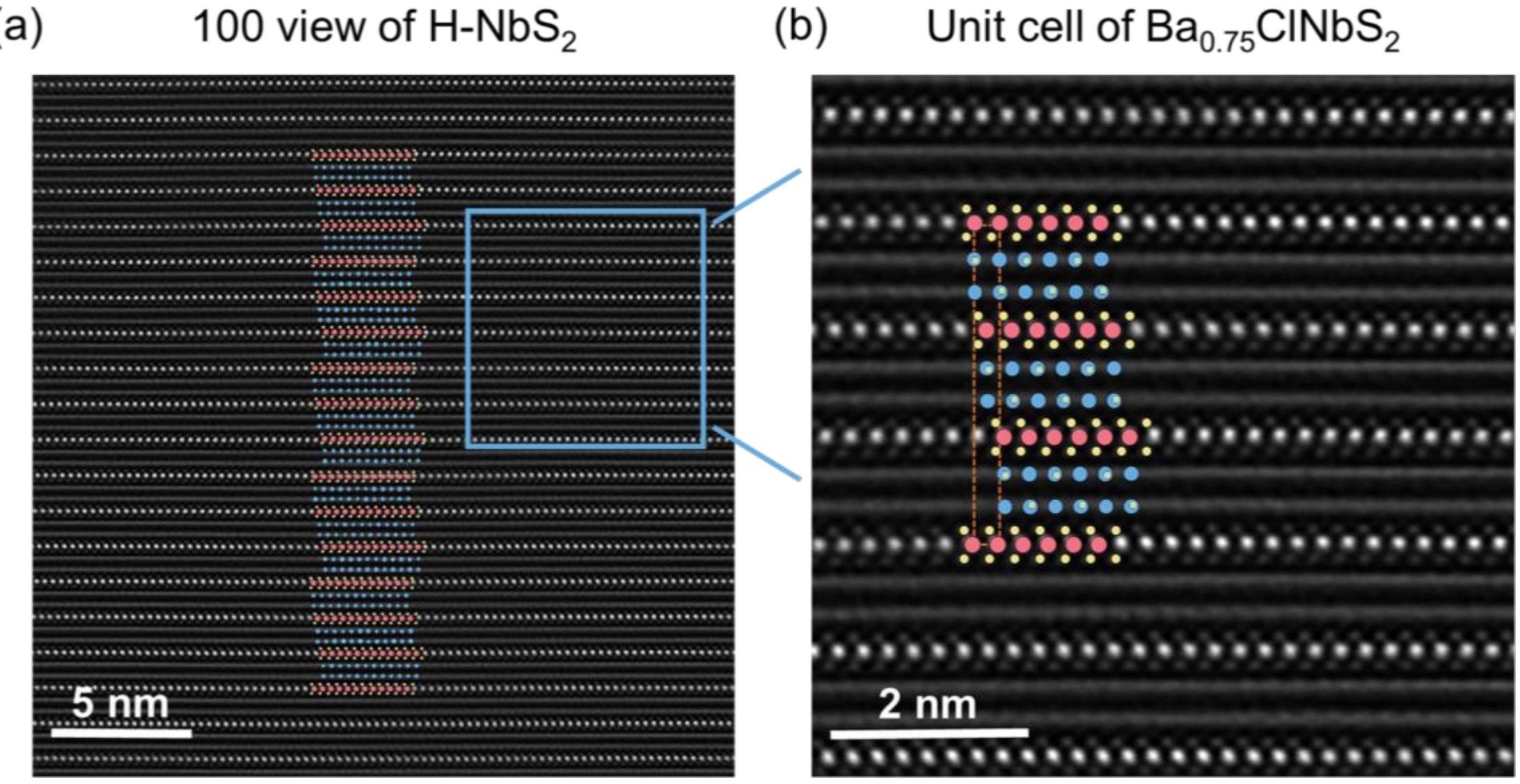

**Fig. S3.** (a,b) Cross-sectional HAADF-STEM overview images of $Ba_{0.75}ClNbS_2$ viewed along the [100] zone axis of the H-$NbS_2$ lattice. The large-area view reveals excellent structural homogeneity across the entire examined region. The 3R primitive unit cells, enclosed by orange squares, stack periodically along the *c*-axis with no observable line defects, stacking faults, or phase segregation.

The atomic structure and stacking sequences of $Ba_{0.75}ClNbS_2$ were unambiguously resolved by cross-sectional high-angle annular dark-field scanning transmission electron microscopy (HAADF-STEM) viewed along the [100] zone axis of the H-$NbS_2$ lattice, with the refined structural model overlaid (Figs. S3(a,b)). Individual H-$NbS_2$ monolayers are clearly separated by Ba-Cl spacer layers, a configuration that intrinsically lacks a spatial inversion center. In contrast to the strictly parallel AA-stacking observed in $Sr_{0.75}ClNbS_2$, the Ba-analogue exhibits a larger out-of-plane lattice parameter accompanied by relative in-plane sliding between adjacent H-$NbS_2$ layers. This structural modulation gives rise to an ABCA stacking sequence characteristic of a rhombohedral 3R phase (Fig. S3b). A large-area STEM image (~ 30 × 30 $nm^2$, Fig. S3(a)) further confirms the highly uniform and periodic arrangement of this ABCA stacking, with the ideal 3R primitive unit cells stacking continuously along the *c*-axis—directly verifying the macroscopic single-phase nature of the synthesized crystals. This well-ordered stacking is likely governed by the specific formation energies associated with Ba-Cl intercalation, which effectively suppresses random layer distribution or intergrowth.

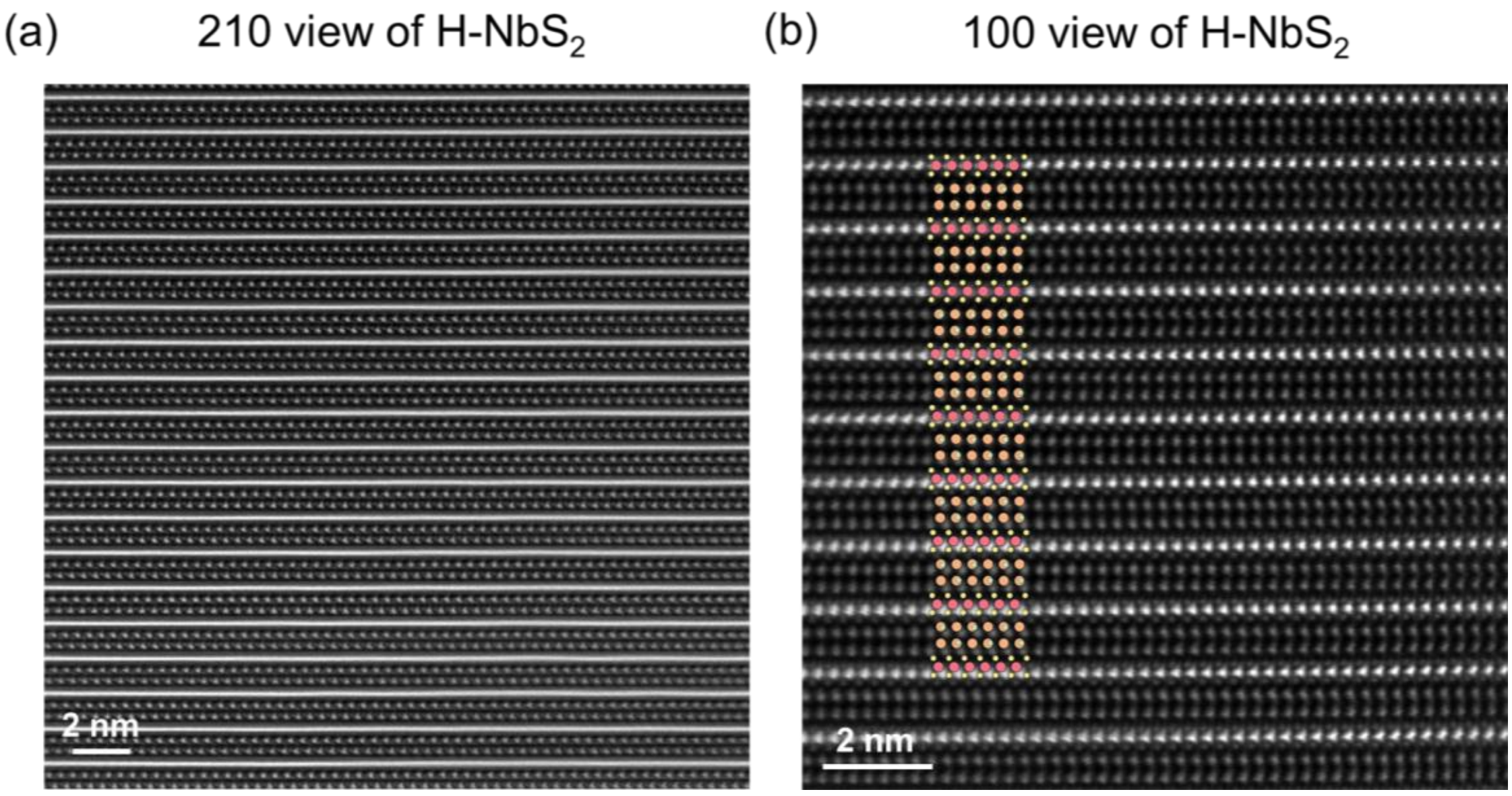

**Fig. S4.** Atomic-scale characterizations of $Sr_{0.75}ClNbS_2$ crystal structure. (a,b) Cross-sectional HAADF-STEM overview images viewed along the [210] and [100] zone axes of the H-$NbS_2$ lattice, respectively, confirming the perfect AA-stacking sequence over large regions without structural defects.

For rigorous comparison, the $Sr_{0.75}ClNbS_2$ crystals were examined under identical conditions (Figs. S4(a,b)). Large-area HAADF-STEM images viewed along both the [210] and [100] zone axes reveal perfectly periodic AA-stacking along the *c*-axis, demonstrating exceptional crystallinity and confirming the complete absence of observable stacking faults or secondary mixed phases. The stark contrast between the 3R-type ABCA stacking in the Ba-phase and the 2H-type AA stacking in the Sr-phase underscores the profound influence of the intercalated cation on the overall stacking configuration.

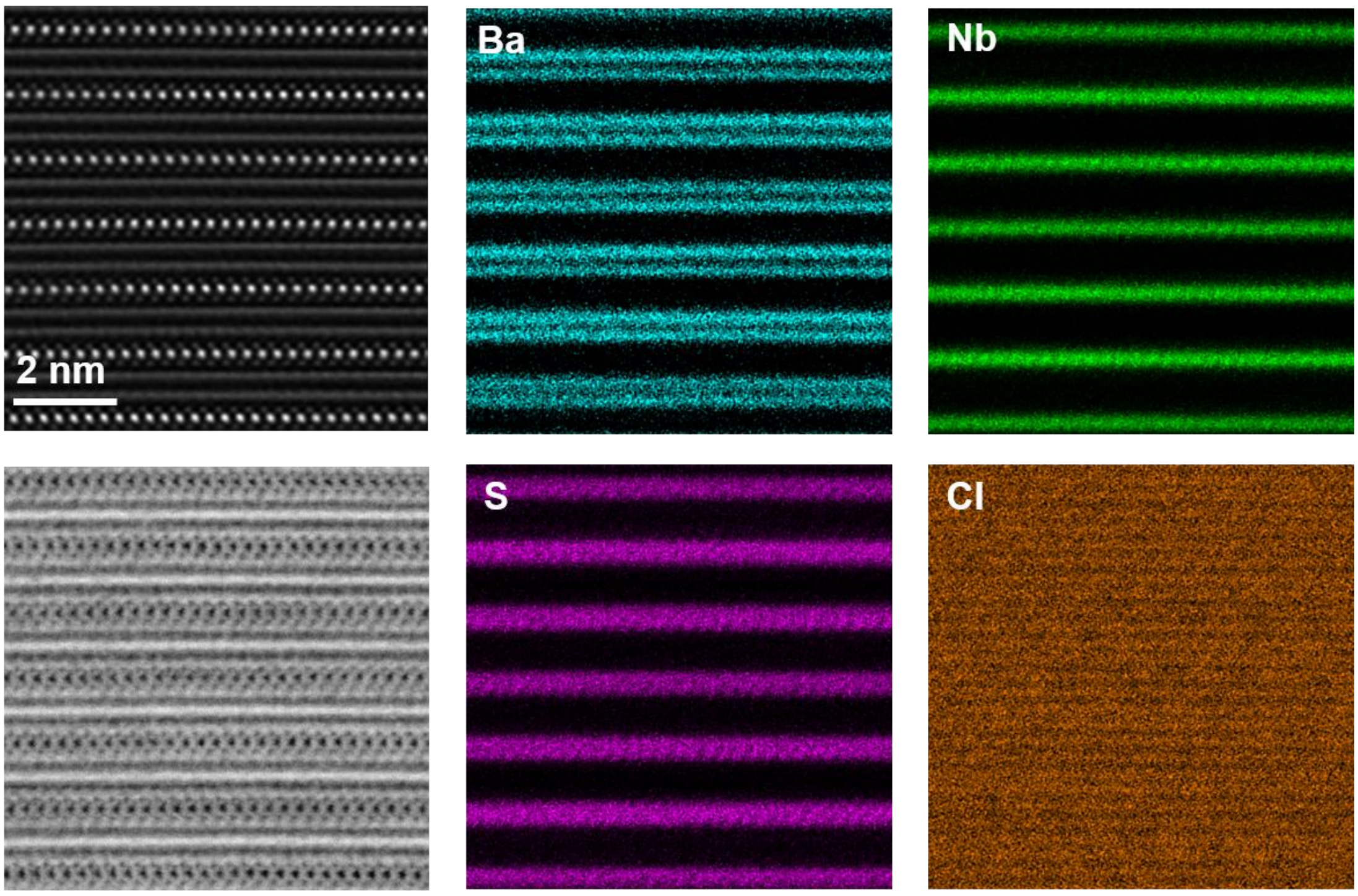

**Fig. S5.** EDS mapping of the $Ba_{0.75}ClNbS_2$ through STEM. Cross-sectional HAADF and bright-field (BF) STEM images of a $Ba_{0.75}ClNbS_2$ single crystal, accompanied by the corresponding atomic-resolution EDS elemental maps.

To unambiguously verify the atomic configurations within the H-$NbS_2$ host layers and the intercalated spacer layers, we performed quantitative atomic-resolution energy-dispersive X-ray spectroscopy (EDS) in conjunction with cross-sectional STEM. The elemental maps (Figs. S5 and S6) reveal a homogeneous distribution of S atoms symmetrically sandwiching the central Nb atomic planes, while Sr (or Ba) atoms are distinctly localized within the van der Waals (vdW) gaps. The Cl elemental mapping, however, appears comparatively indistinct. This reduced clarity may be attributed to limitations in the software-based spectral deconvolution, potentially arising from peak overlaps with adjacent elements or a lower signal-to-noise ratio for the lighter Cl atom, rather than to any genuine non-uniformity in Cl distribution.

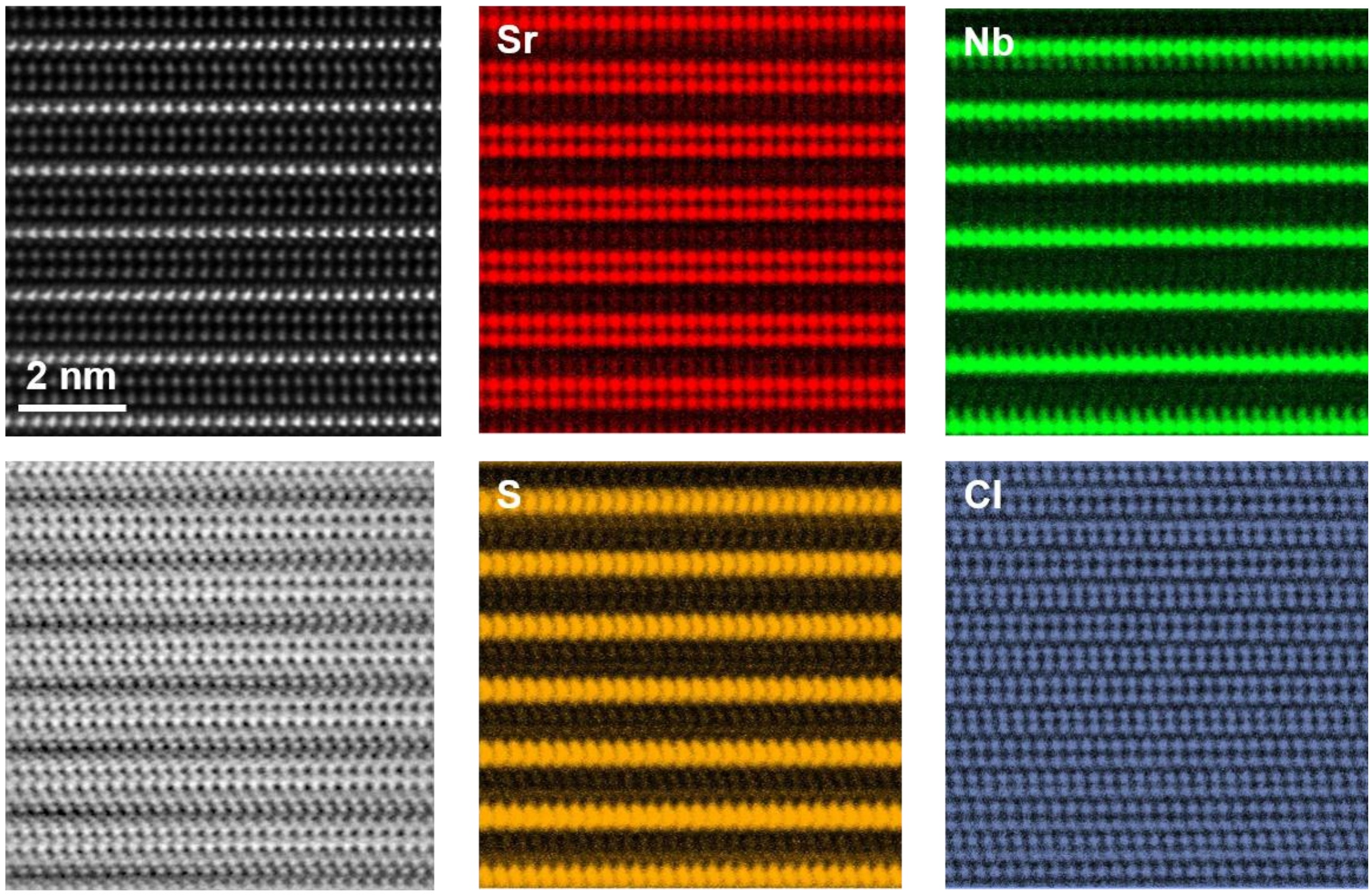

**Fig. S6.** EDS mapping of the $Sr_{0.75}ClNbS_2$ through STEM. Cross-sectional high-angle annular dark-field (HAADF) and bright-field (BF) STEM images of a $Sr_{0.75}ClNbS_2$ single crystal, accompanied by the corresponding atomic-resolution EDS elemental maps.

## IV. Second-harmonic generation (SHG) measurements

All the variable-temperature and polarization-resolved SHG measurements were performed using a home-built optical cryostat operating under high vacuum ($< 1 \times 10^{-8}$ Torr). A femtosecond oscillator (Spectra Physics) was used, whose center wavelength was set to 900 nm and the average power was attenuated to 200 μW, with a pulse duration of ~ 120 fs and a repetition rate of 80 MHz. This excitation beam was focused onto the sample at normal incidence using a 50× microscope objective (numerical aperture, NA = 0.55). The backscattered SHG signal was collected through the same objective, and detected in photon counting mode using a photomultiplier tube (PMT). For the polarization-resolved SHG measurements, SHG signal were acquired by rotating a half-wave plate inserted before the objective. This allowed simultaneous rotation of the polarization angles of both the excitation beam and the detection signal with respect to the sample crystallographic axes. Specific polarization configurations, namely co-polarized (XX) and cross-polarized (XY), were selected.

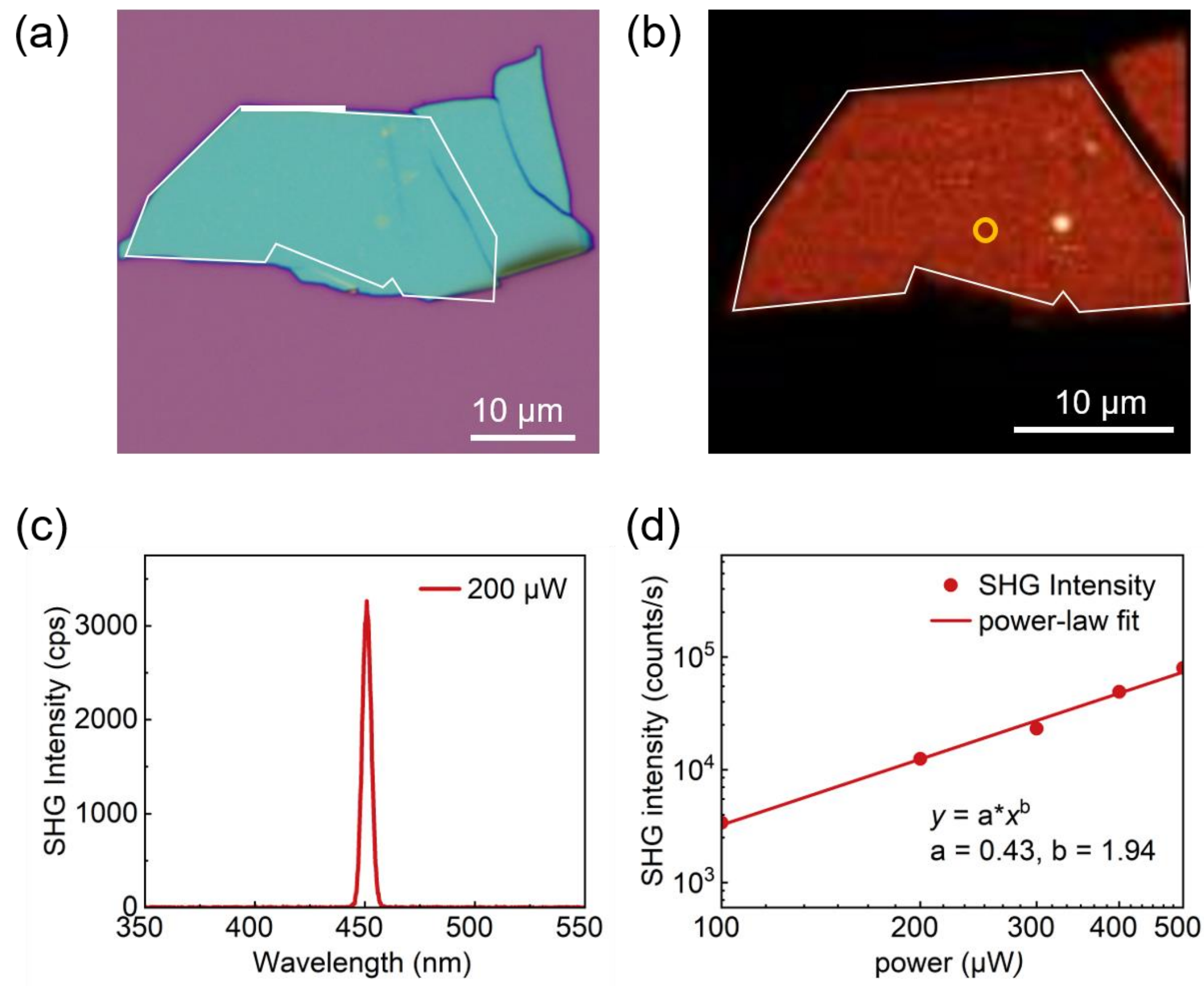

**Fig. S7.** Nonlinear optical SHG characterizations of $Sr_{0.75}ClNbS_2$ single crystals. (a) White-light optical micrograph of a mechanically exfoliated $Sr_{0.75}ClNbS_2$ flake. All subsequent optical data were collected specifically from the localized region indicated by the white circle in (b). (c) The measured SHG intensity spectrum as a function of wavelength. (d) Excitation-power dependence of the SHG signal. The solid line represents a power-law fit to the experimental data, verifying the expected quadratic relationship characteristic of the second-order nonlinear optical process.

## V. Electrical transport measurements

Electrical transport measurements were conducted using a standard four-wire method in a DynaCool Physical Property Measurement System (PPMS-14T) from Quantum Design, equipped with a dilution refrigerator.

As shown in Fig. S8(a), the temperature-dependent longitudinal resistivity ($\rho_{xx}$-$T$) of $Sr_{0.75}ClNbS_2$ and $Ba_{0.75}ClNbS_2$ bulk crystals exhibits an approximately linear relationship at high temperatures, a common signature of non-Fermi-liquid behavior.[1] Below 50 K, the slope of the $\rho_{xx}$-$T$ curve visibly flattens, yielding a moderate residual resistivity ratio ($RRR$ = $\rho_{300K}/\rho_{2K}$) of approximately 5.8 in $Sr_{0.75}ClNbS_2$ and 9.8 in $Ba_{0.75}ClNbS_2$. The insets of Figs. S8(a)and S8(b) highlight the ultralow-temperature superconducting transitions. Notably, $Ba_{0.75}ClNbS_2$ exhibits a severely suppressed onset critical temperature ($T_c$ < 400 mK) and fails to reach a zero-resistance state down to 0.1 K. This drastic suppression of superconductivity, which complicates further transport characterizations, can likely be attributed to the higher concentration of S-site vacancies in the $Ba_{0.75}ClNbS_2$ crystals (as independently confirmed by our EDS analysis in Fig. S1).

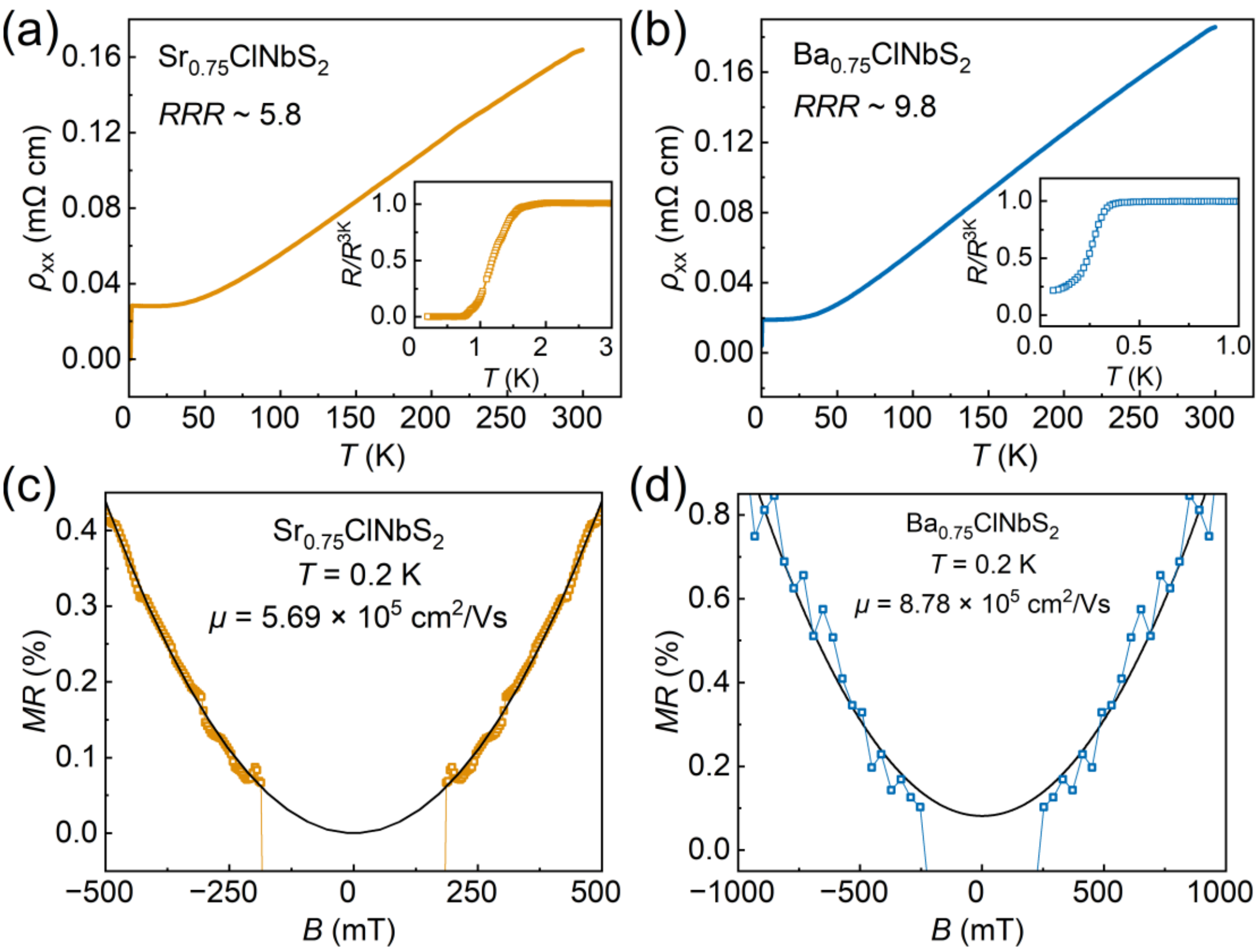

**Fig. S8.** (a,b) Temperature dependence of the longitudinal resistivity for $Sr_{0.75}ClNbS_2$ and $Ba_{0.75}ClNbS_2$, respectively. The insets display the low-temperature resistance normalized to its value

at 3 K, highlighting the superconducting transitions at ultralow temperatures. (c,d) Low-field transverse magnetoresistance measured at 0.2 K. The solid black lines represent the semiclassical quadratic fits used to extract the carrier mobilities.

To assess the electronic quality of the crystals, we extracted the transport mobility from the low-field transverse magnetoresistance (*MR*). In the semiclassical regime, *MR* follows a quadratic magnetic field dependence: $MR = (\rho_{xx} - \rho_0)/\rho_0 = \mu^2B^2$, where $\rho_0$ is the fitted zero-field resistivity in the normal state and $\mu$ is the effective carrier mobility. Figs. S8(c,d) display the low-field MR measured at 0.2 K, along with the corresponding quadratic fits. From these fits, the extracted effective mobility at 0.2 K reaches a remarkably high value of $\mu = 5.67 \times 10^5$ cm² V⁻¹ s⁻¹ for $Sr_{0.75}ClNbS_2$. Such an extraordinarily high carrier mobility reflects a minimal defect-induced carrier scattering rate, fulfilling the stringent prerequisites for realizing and exploring clean-limit two-dimensional superconductivity in this intercalated bulk platform.

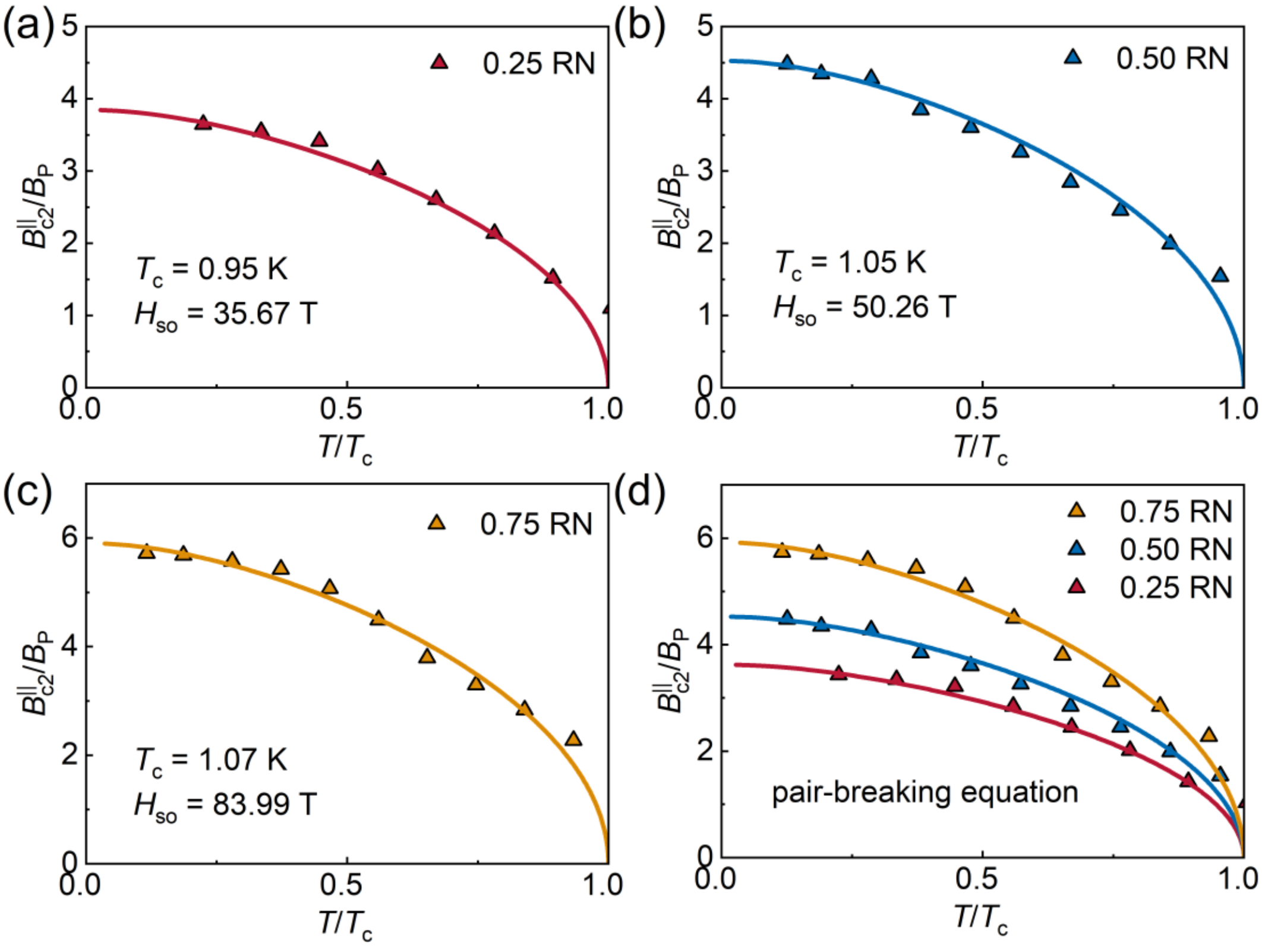

**Fig. S9.** Upper critical fields using different criteria in $Sr_{0.75}ClNbS_2$. (a-d) Temperature dependence

of the in-plane upper critical field $B_{c2}^{||}$ determined using normal-state resistance criteria of 0.25$R_N$, 0.50$R_N$ and 0.75$R_N$. The consistent agreement between the experimental data and the theoretical pair-breaking model across all criteria reliably demonstrates the robust Ising-type superconductivity in $Sr_{0.75}ClNbS_2$.

To rigorously substantiate the presence of Ising-type superconductivity in $Sr_{0.75}ClNbS_2$, we evaluated the in-plane upper critical field $B_{c2}^{||}$ extracted using different normal-state resistance criteria (0.25$R_N$, 0.50$R_N$ and 0.75$R_N$), as presented in Fig. S9. The experimental data across all criteria can be excellently described by the theoretical pair-breaking equation (see Methods), consistently yielding a remarkably high effective Ising spin-orbit field $H_{SO}$ from the corresponding fits[2].

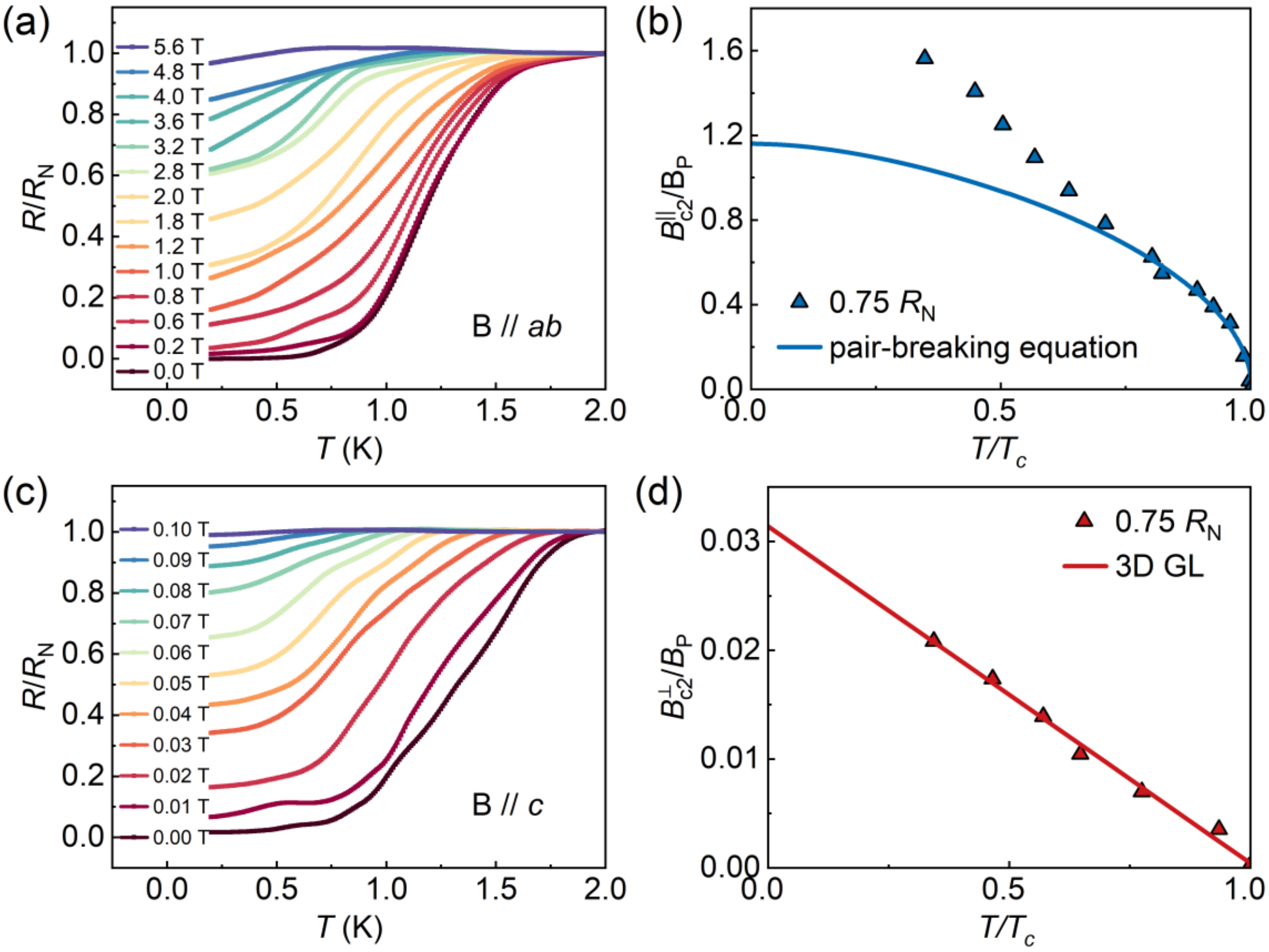

**Fig. S10.** (a) Temperature dependence of the longitudinal resistance measured under various constant in-plane magnetic fields. (b) The corresponding in-plane upper critical field phase diagram fitted using the Ising pair-breaking equation. (c,d)Temperature dependence of the resistance

measured under various perpendicular magnetic fields and the associated upper critical field. The experimental data confirms the dominance of the conventional orbital depairing effect, which is well-fitted by the 3D GL model.

In addition to extracting the in-plane upper critical field from isothermal magnetoresistance sweeps, we systematically measured the temperature dependence of the resistance under various constant in-plane magnetic fields (Fig. S10(a)). The corresponding $B_{c2}^{\parallel}$-$T$ phase diagram (Fig. S10(b)) provides a higher density of data points to precisely capture the depairing behavior near $T_c$. While the overall $B_{c2}^{\parallel}(T)$ relation is well-described by the Ising pair-breaking model, $B_{c2}^{\parallel}$ exhibits a pronounced anomalous upturn at low temperatures, significantly exceeding the Pauli paramagnetic limit. This low-temperature enhancement can be directly attributed to the emergence of the orbital Fulde-Ferrell-Larkin-Ovchinnikov (FFLO) state[3-5]. In the clean-limit, a layered superconductor with the extremely weak interlayer coupling enables finite-momentum Cooper pairs to survive independently within the 2D layers, stabilizing the FFLO phase under massive in-plane magnetic fields.

**VI. ARPES Measurements**

ARPES measurements were performed at the BL03U beamline of the Shanghai Synchrotron Radiation Facility (SSRF), using a Scienta DA30 electron analyzer with vertical slit of 300 μm. The beam spot size is 15 × 15 μm$^2$. The measurements were mainly performed with photon energy of 60 eV. The base pressure during measurements was lower than $1 \times 10^{-10}$ Torr. The measurement temperature was maintained at 10 K. The energy resolution was set to better than 20 meV and angular resolution was about 0.3°. Fig. S11(a) displays the Fermi surface mapping and the corresponding wide-energy-range ARPES spectra of the $Sr_{0.75}ClNbS_2$ single crystals, covering a broad binding energy window from the Fermi level ($E_F$) down to -3.5 eV. We observe a remarkably large energy separation between the bottom of the conduction band and the top of the valence band. This enlarged energy gap is a characteristic signature of the quasi-2D limit, arising directly from the suppression of interlayer interactions that are otherwise strongly mediated by the out-of-plane S $p_z$ orbitals in pristine bulk 2H-$NbS_2$[6-8].

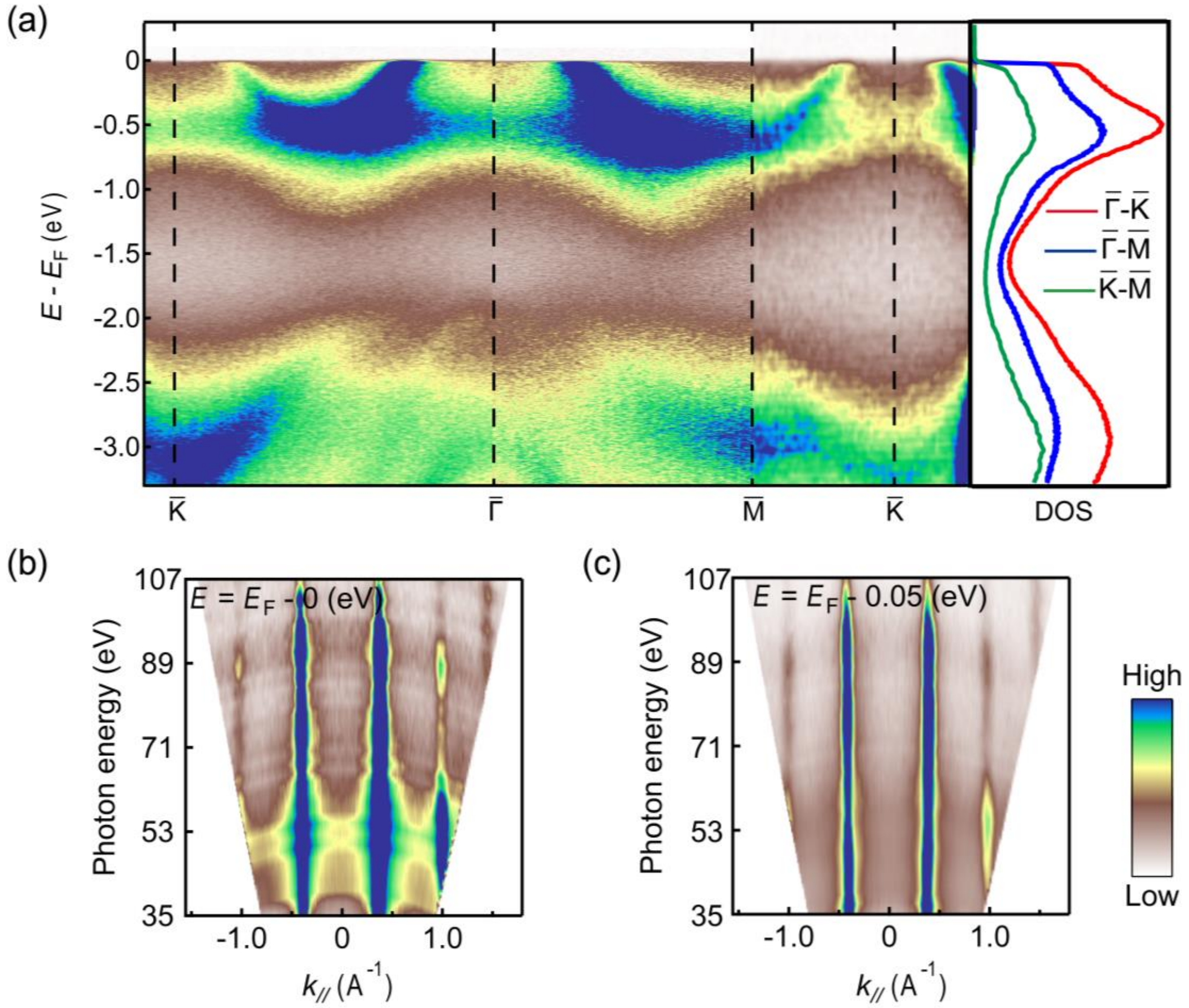

**Fig. S11.** (a) Fermi surface mapping of the $Sr_{0.75}ClNbS_2$ single crystals. (b, c) Photon-energy-dependent ARPES intensity plots illustrating the $k_z$ dispersion of the constant-energy contours at the $E_F$ and a binding energy of 0.05 eV, respectively. The negligible $k_z$ dispersion directly verifies the 2D nature of the electronic states in the bulk crystal.

As discussed in the main text, the in-plane electronic structure closely resembles that of a standalone monolayer. To directly substantiate the 2D nature of these electronic states, we performed photon-energy-dependent ARPES measurements to map the $k_z$ dispersion. Figs. S11(b) and S11(c) visualize the $k_z$-$k_{//}$ constant-energy contours at $E_F$ and a binding energy of 0.05 eV, respectively. The straight and vertical contours reveal a negligible $k_z$ dispersion, which unambiguously confirms the highly 2D nature of the layer-decoupled electronic structure in the macroscopic $Sr_{0.75}ClNbS_2$ bulk crystal.